\documentclass[conference]{IEEEtran}
\IEEEoverridecommandlockouts

\usepackage[english]{babel}
\usepackage{amsmath,amssymb,amsfonts}

\usepackage{colonequals} %
\usepackage{bm} %

\usepackage[
    backend=biber,
    style=ieee,
    sortlocale=en_US,
    sortcites=true,
    url=false,
    eprint=true,
    giveninits=true,
    dashed=false,
    minnames=1,
    maxnames=2
]{biblatex}
\AtEveryBibitem{%
    \clearname{editor}%
    \clearfield{series}%
    \clearfield{isbn}%
    \clearfield{issn}%
    \ifkeyword{keepdoi}
  {}
  {\clearfield{doi}}
    \clearfield{day}%
    \clearfield{month}%
    \clearfield{note}%
    \clearfield{place}%
    \clearfield{addendum}%
    \clearfield{language}
    \clearfield{address}%
    \clearlist{location}%
    \ifentrytype{inproceedings}{%
    }{%
        \clearfield{publisher}%
    }%
    \ifentrytype{software}{%
    }{%
        \clearfield{url}%
        \clearfield{urlyear}%
    }%
}
\usepackage{csquotes} %
\usepackage[inline]{enumitem} %
\usepackage{tabularx} %
\usepackage{booktabs} %
\usepackage[hidelinks]{hyperref} %
\usepackage[capitalize,nameinlink]{cleveref}
\usepackage{xcolor}
\usepackage{graphicx}
\usepackage{braket}
\usepackage{stmaryrd}
\usepackage{mathtools}

\makeatletter
\let\MYcaption\@makecaption
\makeatother
\usepackage{subcaption} %
\makeatletter
\let\@makecaption\MYcaption
\makeatother
\usepackage{wrapfig} %
\usepackage{float}  %
\usepackage{placeins} %
\graphicspath{ {figures/} }
\usepackage{xspace}
\usepackage[ruled,vlined,linesnumbered]{algorithm2e}
\usepackage{fnpct} %
\usepackage{amsthm,thmtools}
\usepackage{orcidlink}
\declaretheoremstyle[
    headformat=\NAME\NUMBER \NOTE,%
    mdframed={},%
    bodyfont=\small%
]{constraintbox}

\crefformat{boxedconstraints}{#2Box C#1#3}
\Crefformat{boxedconstraints}{#2Box C#1#3}

\crefformat{boxedvariables}{#2Box V#1#3}
\Crefformat{boxedvariables}{#2Box V#1#3}

\declaretheoremstyle[
    qed=\textreferencemark
]{example-style}
\declaretheorem[
    name=Example,%
    style=example-style,%
    numbered=unless unique,%
]{example}
\usepackage{siunitx}
\crefname{algocf}{Algorithm}{Algorithms}
\Crefname{algocf}{Algorithm}{Algorithms}
\crefname{section}{Sec.}{Sec.}
\Crefname{section}{Sec.}{Sec.}

\title{Alternating ZX Circuit Extraction
for \\Hardware-Adaptive Compilation\vspace{-0.4cm}}

\author{
    \IEEEauthorblockN{
        Ludwig Schmid\IEEEauthorrefmark{1}$^1$\orcidlink{0000-0002-4246-8125},
        Korbinian Staudacher\IEEEauthorrefmark{2}$^1$\orcidlink{0009-0003-2424-3891},
        and Robert Wille\IEEEauthorrefmark{1}\IEEEauthorrefmark{3}\IEEEauthorrefmark{4}\orcidlink{0000-0002-4993-7860}
    }
    \IEEEauthorblockA{\IEEEauthorrefmark{0}%
      $^1$These authors contributed equally to this work.
    }
    \IEEEauthorblockA{\IEEEauthorrefmark{1}%
    Chair for Design Automation,
        Technical University of Munich, Germany
    }
    \IEEEauthorblockA{\IEEEauthorrefmark{2}%
    MNM-Team,  
    Ludwig-Maximilians-University Munich, Germany,
    }
    \IEEEauthorblockA{\IEEEauthorrefmark{3}%
    Software Competence Center Hagenberg GmbH, Austria
    }
    \IEEEauthorblockA{\IEEEauthorrefmark{4}%
    Munich Quantum Software Company GmbH, Germany
    }

    ludwig.s.schmid@tum.de;
    korbinian.staudacher@nm.ifi.lmu.de;
    robert.wille@tum.de\\
    \url{https://www.cda.cit.tum.de/research/quantum/}
    \vspace{-0.5cm}
}

\begin{document}
    \maketitle

    \begin{abstract}
    We present a novel quantum circuit extraction scheme that tightly integrates graph-like \mbox{ZX diagrams} with hardware-adaptive routing.
    The method utilizes the degrees of freedom during the conversion from a \mbox{ZX diagram} to a quantum circuit  (\emph{extraction}).
    It alternates between generating multiple extraction options and evaluating them based on hardware constraints, allowing the routing algorithm to inform and guide the extraction process.
    This feedback loop extends existing graph-like ZX extraction and supports modular integration of different extraction algorithms, routing strategies, and target hardware, making it a versatile building block during compilation.
    To perform numerical evaluations, a reference instance of the scheme is implemented with SWAP-based routing for neutral atom hardware and evaluated using various benchmark collections on small-~to \mbox{mid-scale} circuits.
    The reference code is available as open-source, allowing fast integration of other extraction and/or routing tools to stimulate further research and foster improvements of the proposed scheme.
    \end{abstract}

    \section{Introduction}\label{sec:intro}

    Executing quantum algorithms on real quantum hardware necessitates several preprocessing steps to translate general quantum operations into an executable, hardware-dependent form.
    As different hardware platforms exhibit different computational capabilities and operational constraints, this requires specialized optimization routines, generally referred to as \emph{compilation}~\cite{chongProgrammingLanguagesCompiler2017,congLightningTalkScaling2023}.
    Essential steps in this process include the generation and optimization of quantum circuits themselves, which we refer to as \emph{circuit synthesis}, and the adaption of the synthesized circuit to a given hardware architecture, including qubit \emph{routing}.
    Typically, these compilation steps are performed independently, with the output of one step serving as the input for the next.
    However, this sequential approach can yield suboptimal results, as the output of one step may not be ideal for the subsequent step~\cite{chongProgrammingLanguagesCompiler2017}.

    For the circuit synthesis step, ZX calculus~\cite{KissingerWetering2024Book} has established itself as a powerful tool for designing~\cite{KissingerWetering2024Book,kissingerPhasefreeZXDiagrams2022,bombinUnifyingFlavorsFault2024} and optimizing~\cite{KissingerWetering2024Book,kissingerReducingNumberNonClifford2020,staudacher2022reducing,vandaeleQubitcountOptimizationUsing2024} quantum circuits.
    It involves converting a quantum circuit into a \mbox{ZX diagram}, simplifying it using diagrammatic rules and, then, converting it back into a quantum circuit through \emph{circuit extraction}~\cite{backens-there-2021,staudacherMulticontrolledPhaseGate2024}.
    This procedure can also be seen as transforming quantum circuits to measurement patterns, performing equivalence-preserving rewrites of the pattern, and extracting the pattern back into a quantum circuit~\cite{duncan-graph-theoretic-2020,waltherExperimentalOnewayQuantum2005}.
    
    Existing circuit optimization algorithms using this pipeline usually operate on a hardware-agnostic level, e.g.,~by optimizing T-gate~\cite{kissingerReducingNumberNonClifford2020} or two-qubit gate count~\cite{staudacher2022reducing,backens-there-2021}. 
    Yet, the best possible circuit may heavily depend on the hardware and the employed routing strategy. 
    In this work, we propose an approach for hardware-adaptive compilation based on the \mbox{ZX calculus}, which iteratively alternates between the circuit extraction and the hardware routing step to guide the extraction process.
    We exploit that the extraction step is not unique and gives rise to a plethora of possibilities, which we call \emph{extraction paths}. 
    These paths are evaluated based on the current hardware configuration, and the best path is selected to continue the extraction.

    \noindent
    Advantages of this approach include:

    \begin{enumerate}
      \item Hardware-adaptive circuit extraction beyond simplistic metrics such as gate count.
      \item Arbitrary hardware platforms can be targeted by employing different routing strategies.
      \item Its simplicity allows for easy integration and adaptation of already existing extraction and routing software, including e.g., future fault-tolerant compilation tools.
    \end{enumerate}

    The general scheme and its functionality are described, and multiple possible extensions to improve the performance of the base version are discussed.
    A reference implementation based on the default ZX extraction algorithm in combination with a SWAP-based routing algorithm for neutral atom hardware is provided.
    This reference implementation is evaluated using random circuits and different benchmarking sets, demonstrating fidelity improvements for the final output circuit.
    The results are discussed, focusing on potential future work.

    The remainder of this paper is structured as follows:
    \Cref{sec:zx} provides a brief review of the \mbox{ZX calculus} and the extraction of graph-like \mbox{ZX diagrams}.
    \Cref{sec:routing} discusses the process of routing circuits to different hardware platforms.
    Then, \Cref{sec:problem} illustrates the issue of suboptimal compilation due to the independent use of extraction and routing algorithms, followed by the proposed scheme in \Cref{sec:solution}.
    In \Cref{sec:eval}, the scheme is evaluated, and the results are compared to the default extraction.
    Finally, we conclude the work in \Cref{sec:conclusion} and discuss future work.

    \section{ZX-Calculus Preliminaries}\label{sec:zx}
    The following section gives a brief overview of the \mbox{ZX calculus} and the extraction of quantum circuits from graph-like \mbox{ZX diagrams}.
    For a comprehensive introduction to the ZX-calculus, we refer to Ref.~\cite{KissingerWetering2024Book}.

    \begin{figure}[t]
        \centering
        \vspace{-24pt}
        \includegraphics[width=0.5\textwidth]{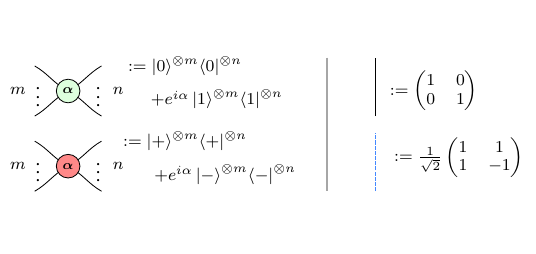}
        \vspace{-1.5cm}
	\caption{Definition of Z(X)-spiders  and the identity (Hadamard) wire.}
	\label{fig:zx-def}
    \end{figure}

    \begin{figure}[t]
    \vspace{-4pt}
        \centering
        \includegraphics[width=0.45\textwidth]{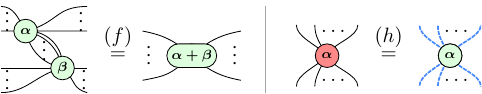}
	\caption{Two of the \mbox{ZX-calculus} rules, namely the fusion rule $(f)$ merging connected spiders and the Hadamard rule $(h)$ inverting colors.}
	\label{fig:zx-rules}
    \end{figure}

    \begin{figure}[t]
        \centering
        \includegraphics[width=0.9\columnwidth]{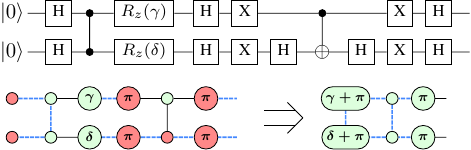}
	\caption{Example of a \mbox{two-qubit} Grover search circuit, which is translated first to a \mbox{ZX diagram} and then converted to a graph-like \mbox{ZX-diagram} using the rules from \Cref{fig:zx-rules}.}
	\label{fig:grover-example}
  \end{figure}

    \subsection{ZX-Calculus}\label{sec:zx-calculus}

    The \mbox{ZX-calculus} is a diagrammatic language for reasoning about linear maps in quantum computing, where nodes (spiders) and edges (wires) form an undirected graph called a ZX diagram.
    There are two types of spiders: green \mbox{\emph{Z-spiders}} and red \mbox{\emph{X-spiders}}.
    Spiders can be parametrized with an angle $\alpha \in [0, 2\pi)$ and correspond to two-dimensional matrices in the Hilbert space with the definition given in \Cref{fig:zx-def}.
    In general, they can have any number of incoming and outgoing wires.
    For convenience, we distinguish between two types of wires: \emph{Normal wires}, representing the identity, and \emph{Hadamard wires}, representing the Hadamard matrix.
    Wires entering the diagram from the left (right) are called \emph{input wires} (\emph{output wires}), with the adjacent spiders defined as \emph{inputs} $I$ (\emph{outputs} $O$).

    \mbox{ZX-diagrams} constitute a graphical language for quantum circuits, meaning any quantum circuit can be represented as a \mbox{ZX-diagram} by replacing gates with equivalent diagrams.
    The power of the \mbox{ZX-calculus} lies in its rules for manipulating these diagrams.
    While not changing the underlying linear map, they can be used to simplify the diagram and, therefore, possibly the resulting quantum circuit.

    Two rules used in the following are shown in \Cref{fig:zx-rules}, namely the fusion rule ($f$) and the Hadamard rule ($h$).
    The fusion rule ($f$) allows merging spiders of the same color if they are connected by at least one normal wire, and ($h$) allows changing the colors of spiders by flipping normal and Hadamard wires.
    All rules apply in both directions and remain valid with interchanged colors.
    These two rules can be extended to form a complete graphical rule set~\cite{vilmart-near-optimal-2018}.

    \begin{example}
    \Cref{fig:grover-example} shows how a \mbox{two-qubit} Grover search circuit is translated to a ZX-diagram, replacing also the \mbox{two-qubit} gates with their equivalent \mbox{ZX-diagram} representation.
    The diagram is then further simplified using the fusion and the Hadamard rule.
    \end{example}

    \subsection{Graph-like \mbox{ZX-diagrams}}\label{sec:zx-graph-like}

    In this work, we focus on the special class of \textit{graph-like} \mbox{ZX-diagrams} as introduced in Ref.~\cite{duncan-graph-theoretic-2020}, which allows us to represent any quantum computation as a graph of parametrized green \mbox{Z-spiders} and Hadamard wires. One can transform any \mbox{ZX-diagram} into an equivalent graph-like \mbox{ZX-diagram} by repeatedly applying standard ZX-rules~\cite{duncan-graph-theoretic-2020}. 
    \begin{example}
    The last step in \Cref{fig:grover-example} shows the graph-like \mbox{ZX-diagram} of the previous \mbox{two-qubit} Grover search circuit.
    \end{example}
    
    Graph-like diagrams can be directly interpreted as patterns in the measurement-based quantum model where each spider corresponds to a qubit in $\ket{+}$ state and each Hadamard wire to a $CZ$ gate.
    As such, they provide additional transformations such as \emph{local complementation}, \emph{pivoting}, and \emph{phase gadget elimination}, where the latter was introduced in Ref.~\cite{kissingerReducingNumberNonClifford2020}.
    Using these rules to simplify graph-like diagrams has become a default procedure and is offered by tools such as \texttt{PyZX}~\cite{kissinger2020Pyzx}.
    \subsection{Circuit Extraction}\label{sec:zx-circuit-extraction}

    \begin{figure}[t]
    \includegraphics[width=0.45\textwidth]{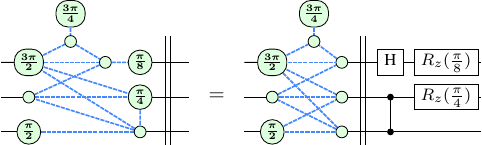}
	\caption{Basic circuit extraction from a graph-like ZX diagram.}
	\label{fig:zx-extraction}
    \end{figure}

    \begin{figure*}[ht]
    \vspace{-0.6cm}
        \includegraphics[width=0.9\textwidth]{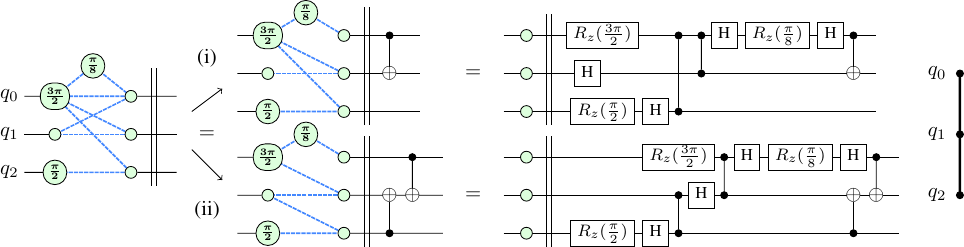}
	\caption{Full ZX circuit extraction of a graph-like diagram using a CX gate to rearrange the graph connections.}
	\label{fig:zx-extraction-cnot}
    \end{figure*}

    While the diagrammatic language is a powerful tool for reasoning and optimizing quantum operations, as a final step, to obtain a quantum circuit with the same number of qubits as the original, one can \emph{extract} a quantum circuit from the ZX diagram.
    This extraction is feasible in polynomial time if the underlying graph has some kind of \emph{flow}~\cite{backens-there-2021,simmons2021}.
    Here, we provide a brief overview of the extraction algorithm for graph-like \mbox{ZX diagrams} with gflow as described in Ref.~\cite{backens-there-2021}.

    The algorithm extracts a quantum circuit by identifying suitable parts of the \mbox{ZX diagram} and converting them into their equivalent quantum gates.
    These parts are then removed from the diagram, extracting one gate at a time until only the inputs and outputs remain.
    During the process, a set of green \mbox{Z-spiders} called the \textit{frontier} separates the extracted portion of the diagram from the unextracted part.

    First, there are three basic extraction rules:
    \begin{itemize}
        \item \textbf{Phase:} Phases of frontier spiders are directly extracted as $R_z$ gates.
        \item \textbf{CZ:} Hadamard wires between frontier spiders are extracted as CZ gates.
        \item \textbf{Hadamard:} Hadamard wires where a frontier spider is only connected to a \mbox{non-frontier} spider are extracted as Hadamard gates.
    \end{itemize}

    \begin{example}
    \Cref{fig:zx-extraction} illustrates the three above rules, extracting four gates from the diagram.
    The double vertical lines visualize the separation between the diagram and the circuit.
    \end{example}

    Second, if every spider in the frontier has at least two \mbox{non-frontier} neighbors, one can add wires of a frontier spider to another by placing a CX gate in the extracted circuit with the following effect:
    \begin{enumerate}
        \item \textbf{CX:} Extracting a CX gate with control $c$ and target \mbox{qubit $t$} copies all Hadamard wires of the target frontier spider to the control qubit frontier spider. Double wires cancel each other.
    \end{enumerate}

    \begin{example}\label{ex:zx-extraction-cnot}
    In the upper part (i) of \Cref{fig:zx-extraction-cnot}, a CX gate between control qubit $q_0$ and target $q_1$ is extracted.
    The two Hadamard wires of the frontier spider $q_1$ are copied to $q_0$ and cancel the existing Hadamard wires.
    After this CX gate, the rest of the diagram can be extracted, as shown on the right.
    \end{example}

    Previous work focused on finding optimized CX sequences and graph manipulations to reduce hardware requirements, such as \mbox{two-qubit} count~\cite{staudacher2022reducing,backens-there-2021}.
    In addition, there exist special-purpose extraction algorithms, such as for \mbox{multi-controlled} phase gates~\cite{staudacherMulticontrolledPhaseGate2024}.
    In the following, we discuss how such extracted circuits can then be executed on hardware.

    \section{Hardware Mapping \& Routing}\label{sec:routing}
    While the \mbox{ZX calculus} is a powerful tool to construct optimized quantum circuits, referred to as circuit \emph{synthesis}, to execute these circuits on real hardware, further compilation steps are required~\cite{chongProgrammingLanguagesCompiler2017,congLightningTalkScaling2023}.
    First \emph{mapping}, which corresponds to a bijective assignment between the circuit qubits and the available hardware qubits.
    Second \emph{routing}, where the limited hardware qubit connectivity is circumvented by inserting additional operations.
    Depending on the hardware platform and the exact hardware specification, this step can be realized differently.
    This encompasses the insertion of SWAP gates~\cite{zulehnerEfficientMethodologyMapping2019,cowtanQubitRoutingProblem2019,tanOptimalLayoutSynthesis2020,liTacklingQubitMapping2019}, shuttling operations on QCCD \mbox{trapped-ion} architectures~\cite{schoenbergerShuttlingScalableTrappedIon2024a,muraliArchitectingNoisyIntermediateScale2020,kreppelQuantumCircuitCompiler2023,sakiMuzzleShuttleEfficient2022}, or atom reconfigurations on the neutral atom platform~\cite{tanCompilingQuantumCircuits2023,linReuseAwareCompilationZoned2024,stadeAbstractModelEfficient2024, schmidHybridCircuitMapping2023,ludmirParallaxCompilerNeutral2024}.
    A comprehensive discussion of the different methods and the corresponding software packages is out of the scope of this work, but the following are some general aspects of hardware compilation:
    \begin{enumerate}[leftmargin=*, label={}, itemindent=0pt, labelsep=0pt]
    \item \textbf{First,} typically, the mapping and routing steps are executed after circuit synthesis.
      This is based on the promise that optimizing each step independently will yield a good overall result while facilitating the development of specialized algorithms for each step.
  \item \textbf{Second}, optimization methods typically focus on some metric or proxy, such as (\mbox{two-qubit}) gate count. 
    Yet, in all metrics, the overall cost function is the infidelity of the output circuit, which we aim to minimize in the mapping and routing process.
    \item \textbf{Third}, the diversity of existing hardware, even within the same platform, requires highly specialized compilation methods to make optimal use of the available hardware.
    As a result, often only a small subset of the available compilation methods can be employed for a given hardware setup.
  \end{enumerate}
    Within this work, we model the hardware compilation process as a black box, which takes a quantum circuit and a hardware architecture as inputs and returns a mapped quantum circuit together with some cost metrics indicating the overhead introduced by the mapping and routing process.

    \section{Considered Problem}\label{sec:problem}

    In the following, we illustrate how the disjoint use of ZX-based circuit synthesis and hardware routing can lead to suboptimal results.
    For that, let us revisit the circuit extraction from \Cref{ex:zx-extraction-cnot} shown in the upper part (i) of \Cref{fig:zx-extraction-cnot}.

    The circuit was extracted using a single additional CX gate, resulting in a total of three entangling gates (2 CZ and 1 CX) requiring interaction between qubits $q_0 \leftrightarrow q_2$ and $q_0 \leftrightarrow q_1$.

    But, as discussed in \Cref{sec:zx-circuit-extraction}, this extraction step is not unique, and there is a certain degree of freedom in the extraction process~\cite{duncan-graph-theoretic-2020}.
    Alternatively, one could have inserted a second CX gate between $q_2$ and $q_3$ to further reduce the number of Hadamard wires in the graph.
    This step and its result are shown as (ii) below, where the resulting circuit has four \mbox{two-qubit} gates (2 CZ and 2 CX) and requires connectivity between qubits $q_0 \leftrightarrow q_1$ and $q_1 \leftrightarrow q_2$.

    For circuit extraction, a typical metric is the number of entangling gates, as they are a major source of error on available hardware~\cite{preskillQuantumComputingNISQ2018}.
    Thus, the first circuit (i) would be preferable as it requires three entangling gates compared to four in the second circuit (ii).

    However, this changes if one simultaneously considers the necessary routing.
    As an example, we assume SWAP-based routing and a linear hardware configuration as given on the very right of \Cref{fig:zx-extraction-cnot}.
    While the second circuit can be executed directly on the hardware, as all required qubit connections ($q_0 \leftrightarrow q_1$ and $q_1 \leftrightarrow q_2$) are directly available, the first circuit contains an unavailable \mbox{two-qubit} gate connection ($q_0 \leftrightarrow q_2$).
    The necessary SWAP insertion between $q_0$ and $q_1$ would result in an additional 3 CX gates, resulting in $3+3=6$ \mbox{two-qubit} gates for the first circuit (i). 
    This makes the second circuit (ii) the better choice with its four \mbox{two-qubit} gates, as it inherently matches the hardware connectivity.

    This simple example illustrates that a circuit that might be preferable based on metrics, such as the number of \mbox{two-qubit} gates, eventually can result in more costly routing and, therefore, in suboptimal final output.
    Although the example only considered SWAP gate insertions, the idea also applies to other routing strategies, such as shuttling or atom reconfigurations.

    To circumvent this suboptimal interplay, we propose an alternating scheme where the extraction and routing algorithms communicate iteratively to avoid compilation steps that might be suboptimal for the opposite party.

    \section{Alternating Circuit Extraction}\label{sec:solution}
    In this section, we present a solution to address the issue of suboptimal compilation results arising from the disjoint use of ZX circuit extraction and hardware routing.
    First, we give a high-level overview of the proposed solution, followed by a detailed description of the implementation and further refinements.
    Third, we briefly discuss a concrete instance to realize this general scheme in practice, which will be used in the evaluations of \Cref{sec:eval}.

    \subsection{Overview}\label{sec:solution-overview}
    \begin{figure}[t]
    \centering
        \vspace{-0.2cm}\includegraphics[width=0.95\columnwidth]{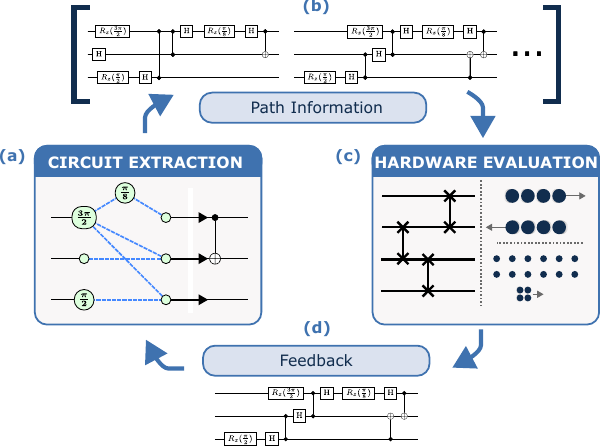}
        \vspace{-0.2cm}
	\caption{Overview of the cross-compilation strategy and the communication between circuit extraction and routing.\vspace{-0.2cm}}

	\label{fig:cross-compilation}
    \end{figure}
    We propose a circuit extraction scheme alternating between the extraction of graph-like \mbox{ZX diagrams} and the hardware routing, which satisfies the given hardware constraints.
    The abstract idea is illustrated in \Cref{fig:cross-compilation}, revisiting the circuits from the previous examples.
    The extraction procedure consists of the following steps:

    \begin{enumerate}[leftmargin=*,label=\textbf{(\alph*)}]
      \item \textbf{Circuit Extraction:} Within the ZX extraction, different possible extraction paths are generated with Phase, CZ, Hadamard and CX extraction each applied once.
      \item \textbf{Path Information:} Descriptions of the paths (their full circuit or certain metrics) are sent to the routing process.
      \item \textbf{Hardware Evaluation:} The routing algorithm evaluates the paths based on the current hardware configuration and selects the best path based on a chosen metric.
      \item \textbf{Feedback:} The selection is sent back to the extraction algorithm, which continues the extraction process along the selected path.
    \end{enumerate}
    Through this iterative back-and-forth communication, the entire graph is converted into a fully mapped quantum circuit.
    It should be noted that this scheme is quite general and not limited to a particular extraction algorithm, evaluation metrics, routing strategy, or hardware platform.
    In particular, any of the extraction algorithms from \Cref{sec:zx-circuit-extraction} and any hardware compilation method from \Cref{sec:routing} could be employed.
    Furthermore, this could also be generalized to fault-tolerant compilation methods in a similar way.

    While the feedback allows the extraction algorithm to make better-informed decisions, it also imposes an additional computational overhead for the iterative path evaluations.
    Let~$k$ denote the number of spiders in the diagram and $m$ the number of paths at each extraction step. Since at each extraction step we remove at least one spider, we get an upper bound of $\mathcal{O}(km)$ evaluations of the routing cost.
    Assuming $k$ to be constant for a given diagram, this results in a typical search-space exploration \mbox{trade-off}.
    By considering more extraction options, one will likely improve the output quality of the scheme while simultaneously increasing the necessary computational resources.
    Finding efficient heuristics for search space exploration and pruning constitutes, therefore, an essential task for future research to scale the scheme to larger systems.
    One should note that, given the resources, the costly evaluation of different extraction paths can be done in parallel.

    \subsection{Improving the Base Scheme}\label{sec:improvements}
    Based on the above scheme, we introduce several possible modifications that can further improve the performance of the base version.
    The different modifications are then also evaluated in \Cref{sec:eval}.

    \noindent
    \textbf{Edge Bias Weight $\beta$:}
    While all extraction paths are valid options, they are not necessarily equivalent regarding the ``amount of graph'' that is extracted.
    Simultaneously, shorter circuits are typically easier to route.
    As a result, the hardware routing might prefer shorter circuits over paths, which would result in a faster extraction.
    To account for this, we introduce a cost bias with weight $\beta$ to the evaluation metric, namely
    \begin{equation}
      C_\mathrm{bias} = \beta \cdot \Delta W \,,
        \label{eq:edge-bias}
    \end{equation}
    where $\Delta W$ is the reduction of Hadamard wires in the graph for a certain extraction path.
    Increasing $\beta$ will favor extracting larger parts of the graph while potentially overshadowing the actual routing cost.

    \noindent
    \textbf{Sliding Window $s$:}
    Many routing algorithms make heavy use of contextual information, including a certain lookahead to find globally favorable routing operations~\cite{cowtanQubitRoutingProblem2019,zulehnerEfficientMethodologyMapping2019,schmidHybridCircuitMapping2023}.
    Considering every extraction circuit by itself neglects the surrounding circuit, resulting in locally optimized but globally suboptimal routing.
    This drawback affects both the evaluations of the different paths and the routing of the final output.

    To circumvent this, we introduce a sliding window approach.
    Given the size of the window $s$, the routing algorithm will consider not only the current path but also the $s$ previously extracted gates.
    For $s$ sufficiently large, this results in a complete rerouting of the full extracted circuit.
    While giving more context to the routing process, it introduces additional overhead for each of the $\mathcal{O}(km)$ path evaluations.

    \noindent
    \textbf{Extraction Depth $l$:}
    To provide even more context to the routing algorithm, one can also increase the extraction \mbox{depth $l$} which determines how often we run a cycle of Phase, Hadamard, CZ and CX extraction before hardware evaluation.
    This will provide the routing algorithm with all possible paths for $l$ consecutive extractions.
    While providing more information to the routing algorithm, this exponentially increases the number of paths to be evaluated to $\mathcal{O}((km)^l)$, making it practically feasible only for small values of $l$.

       \section{Evaluations}\label{sec:eval}

    While the scheme itself is general, we need to implement a concrete instance for numerical evaluations.
    For each of the parts in \Cref{fig:cross-compilation}, we use the following tools:

    \noindent
    \textbf{(a + b)} For the ZX extraction phase, we modify the ``default'' extraction algorithm as described in Ref.~\cite{backens-there-2021} and send the complete circuits to the routing algorithm.
    For the representation of the graph-like \mbox{ZX diagrams}, we use \texttt{PyZX}~\cite{kissinger2020Pyzx}.

    \noindent
    \textbf{(c + d)} Hardware routing is realized using the neutral atom hybrid routing~\cite{schmidHybridCircuitMapping2023} from the Munich Quantum Toolkit (MQT~\cite{willeMQTHandbookSummary2024}). 
    While developed for neutral atom hardware, we only focus on the SWAP insertion part using the SABRE method~\cite{cowtanQubitRoutingProblem2019}, which could also be used for superconducting hardware.
    The best path is selected based on the \emph{approximate success probability} (ASP), as defined in Ref.~\cite{schmidComputationalCapabilitiesCompiler2024}
    \begin{equation}
        \text{ASP} = \exp \left( -\frac{t_\mathrm{idle}}{T_\mathrm{eff}} \right) \prod_{i=1}^{N} \mathcal{F}(O_i) \, ,
        \label{eq:asp}
    \end{equation}
    where $\mathcal{F}(O_i)$ represents the fidelity of the $i$-th operation, $t_\mathrm{idle}$ is the idle time, and $T_\mathrm{eff} = \frac{T_1 T_2}{T_1 + T_2}$ denotes the effective coherence time.
    The ASP functions as a proxy to the actual fidelity expected on hardware by considering hardware parameters such as coherence times and gate fidelities.
    
    The complete code of this reference implementation is available as open-source on Zenodo~\cite{schmidAlternatingZXCircuit2026}. %

    \subsection{Evaluation Setup} 
    To study the performance and behavior of the proposed scheme, we perform numerical simulations using the setup described above.
    First, the influence of the different parameters, such as the extensions of \Cref{sec:improvements}, are evaluated.
    Based on this, we apply the scheme to different benchmarking sets.
    Throughout this section, all circuits are converted to \mbox{ZX diagrams} and optimized using the \texttt{PyZX} \texttt{full\_reduce} optimization.
    As a baseline, we employ the default \texttt{PyZX} extraction algorithm \texttt{extract\_circuit} to generate the corresponding quantum circuit.
    We consider the ASP from \Cref{eq:asp} of the final routed circuit as the target metric and calculate the relative difference compared to the baseline, referred to as $\Delta$ Fidelity.
    The hardware configuration is a $6\times6$ nearest neighbor grid with parameters taken from \cite{schmidComputationalCapabilitiesCompiler2024}. More details, including all evaluation scripts, data, and figures, are available on Zenodo~\cite{schmidAlternatingZXCircuit2026}.

    \subsection{Random Circuits}\label{sec:random-circuits}

    \begin{figure}[t]
    \centering
    \vspace{-0.25cm}
        \includegraphics[width=0.95\columnwidth]{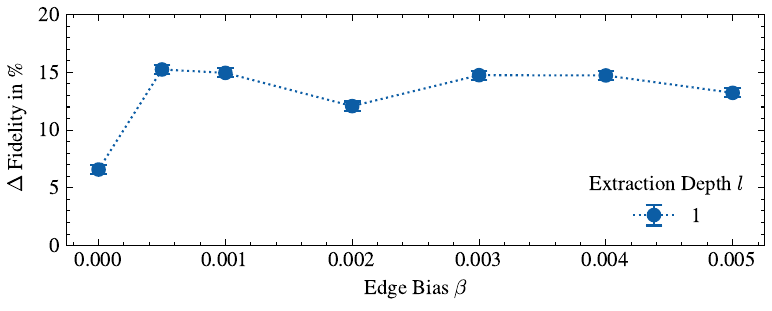}
        \vspace{-0.25cm}
	\caption{Change of fidelity compared to default extraction for different edge weight biases $\beta$ on random Clifford+T circuits.}
        \label{fig:random_bias}
    \end{figure}

    \begin{figure}[t]
    \centering
    \vspace{-0.2cm}
        \includegraphics[width=0.95\columnwidth]{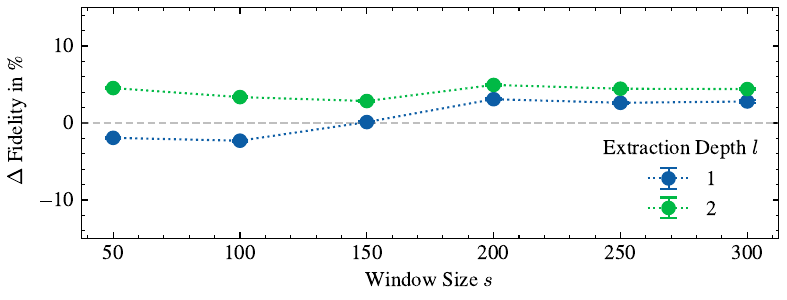}
        \vspace{-0.25cm}
	\caption{Change of fidelity compared to default extraction for different edge window sizes~$s$ and extraction depths~$l$ on random Clifford+T circuits.}
        \label{fig:random_steps}
    \end{figure}

    \begin{figure*}[t]
    \centering
    \vspace{-1.1cm}
        \includegraphics[width=1\textwidth]{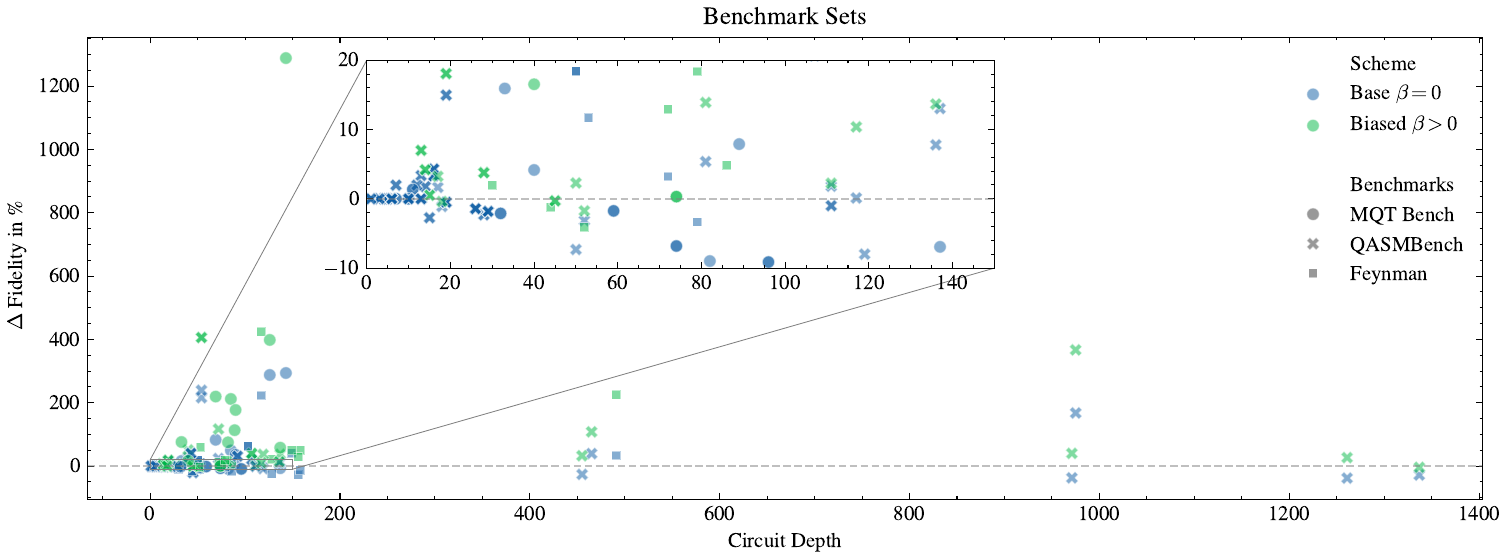}
        \vspace{-0.75cm}
	\caption{Change of fidelity compared to the default extraction for different benchmark sets. The base scheme (blue) refers to no edge bias $\beta=0$, while the biased scheme (green) was chosen over a range of edge biases.}
        \label{fig:benchmarks}
    \end{figure*}

    We consider random Clifford+T circuits generated using \texttt{PyZX} \texttt{randomT} function with $n=6$ qubits and a circuit depth of $d=300$.
    The T~(CX) gate probability is set to $p_\mathrm{T} = 0.4$ ($\,p_\mathrm{CX} = 0.3$).

    First, \Cref{fig:random_bias} shows the result of varying edge weight bias~$\beta$ with the sliding window size set to the full circuit depth.
    For the considered circuits, the proposed scheme always improves the ASP of the output circuit.
    The data shows how the fidelity improvement increases for nonzero edge biases~$\beta$, demonstrating the positive effect of including the number of extracted Hadamard wires into the cost function.
    For even larger biases, the performance drops slightly again, which may be due to the fact that the bias overshadows the actual routing cost, as discussed in \Cref{sec:improvements}.

    Second, \Cref{fig:random_steps} shows the results of the other two introduced parameters, namely the varying window size~$s$ and extraction depth~$l$.
    As expected, it shows a performance increase for larger window sizes, supporting the hypothesis that providing more context to the routing algorithm is beneficial.
    It saturates at about $s \approx 2/3\, d$ of the total circuit depth.
    Similar output quality can be achieved by increasing the extraction depth $l$, which also facilitates the discovery of good extractions by providing more and deeper extraction information.
    Here, an edge bias of~$\beta=0.002$ was chosen.

    \subsection{Benchmarking Circuits}\label{sec:benchmarks}
    
    We evaluate circuits with qubit numbers ranging from $2$~to $14$ and circuit depths from $1$ to $1337$.
    The circuits are taken from common benchmarking libraries, including MQT Bench~\cite{quetschlichMQTBenchBenchmarking2023}, QASM Bench~\cite{liQASMBenchLowlevelQASM2022}, and the Feynman benchmarking set~\cite{amyLargescaleFunctionalVerification2019}.
    \Cref{fig:benchmarks} shows the relative difference of the ASP metric for different benchmarking sets, distinguished by different markers.
    The colors distinguish between the base scheme with $\beta=0$ and a biased version, where the edge weight was optimized over different values $\beta \in \{5\times 10^{-5},\ 5\times 10^{-4},\ 1\times 10^{-3},\ 3\times 10^{-3},\ 5\times 10^{-3},\ 0.01\}$.
    The extraction depth was set to $l=1$, and the window size was set to $s=\infty$, completely rerouting the whole circuit.

    While for shallow circuits (typically preparation of certain states) the performance is similar to the default extraction, for deeper circuits, there can be large fidelity improvements of up to 250\% for the biased version.
    At the same time, there also exist a few circuits where the default \texttt{PyZX} is slightly superior, even to the biased version.
    These circuits seem to have a specific structure that does not allow the routing evaluation to choose an overall beneficial extraction path.
    Future research should investigate this further and how to potentially circumvent this issue.

    \subsection{Discussion}\label{sec:discussion}
    The proposed setup performs well for the considered small and mid-sized circuits.
    However, as the employed routing algorithm relies heavily on the context of the circuit, it requires repetitively rerouting large parts of the extracted circuit.
    In combination with the edge weight bias, the method requires specific hyperparameter optimization to outperform the default \texttt{PyZX} extraction.
    Nevertheless, the proposed scheme formulates a promising approach to realize hardware-adaptive ZX extraction.
    First, it extends the default ZX extraction for hardware-adaptive compilation beyond simple gate count metrics, where the original procedure is recovered if one considers only a single extraction path.
    Second, the structure is modular, where both extraction and routing can be replaced by other algorithms. 
    This allows the scheme to be applied to arbitrary hardware platforms and their unique specialties while using existing compilation software.
    Third, its simplicity allows for fast integration of upcoming routing methods and adapting existing tools.

    \section{Conclusion \& Outlook}\label{sec:conclusion}
    We introduced an alternating scheme for quantum circuit extraction that combines \mbox{ZX diagram} extraction with \mbox{hardware-adaptive} routing through continuous feedback.
    By evaluating multiple extraction paths against hardware constraints and routing costs, the method enables better-informed extraction decisions and extends existing ZX extraction techniques.
    The modular structure allows for integration with various extraction algorithms and routing tools, making it applicable across different hardware architectures.
    It works on top of existing routing and compilation software, providing further improvements with little to no modifications.

    This initial work opens several avenues for future research.
    First, one can consider additional extraction steps, for instance, the extraction of \mbox{multi-controlled} phase gates~\cite{staudacherMulticontrolledPhaseGate2024}.
    Second, the extension to alternative routing software, e.g., for trapped ion systems or even fault-tolerant routing.
    Finally, by investigating improved heuristics and search space pruning techniques, one should be able to reduce the number of paths to be evaluated while maintaining similar output quality.

    \section*{Acknowledgments}
    
    This work is partially supported by the German Federal Ministry of Education and Research (BMBF) under the funding program Quantum Technologies - From Basic Research to Market under contract number 13N16070.
    The authors acknowledge funding from the Munich Quantum Valley initiative (K5), which is supported by the Bavarian state government with funds from the Hightech Agenda Bayern Plus.
    
    L.S. and R.W. acknowledge funding from the European Research Council (ERC) under the European Union's Horizon 2020 research and innovation program (Grant Agreement No. 101001318 and No. 101114305 -“MILLENION-SGA1”).
    Furthermore, this work was supported by the BMFTR under grant number 13N17298 (SYNQ) and the Deutsche Forschungsgemeinschaft (DFG, German Research Foundation) under grant numbers 563402549 and 563436708.

    \renewcommand*{\bibfont}{\small} %
    \printbibliography
    
\end{document}